\documentclass[preprint,12pt, a4paper]{elsarticle}

\usepackage{amsmath}
\usepackage{amssymb}
\usepackage{array}
\usepackage{booktabs}
\usepackage{color}
\usepackage{float}
\usepackage{graphicx}
\usepackage{ifpdf}
\usepackage[utf8]{inputenc}
\usepackage{keyval}
\usepackage{lineno}
\usepackage{listings}
\usepackage{moresize}
\usepackage{multirow}
\usepackage{paralist}
\usepackage{rotating}
\usepackage{soul}
\usepackage{srcltx}
\usepackage{url}
\usepackage{xcolor}
\usepackage{xspace}
\usepackage{wrapfig}
\usepackage{subfig}
\usepackage[export]{adjustbox} 

\definecolor{listinggray}{gray}{0.95}
\definecolor{darkgray}{gray}{0.7}
\definecolor{commentgreen}{rgb}{0, 0.4, 0}
\definecolor{darkblue}{rgb}{0, 0, 0.6}
\definecolor{purple}{rgb}{0.6, 0, 0.6}
\definecolor{middleblue}{rgb}{0, 0, 0.75}
\definecolor{darkred}{rgb}{0.4, 0, 0}
\definecolor{brown}{rgb}{0.5, 0.5, 0}
\definecolor{dkgreen}{rgb}{0,0.5,0}
\definecolor{orange}{rgb}{1,.5,0}
\definecolor{dandelion}{cmyk}{0,0.29,0.84,0}

\usepackage[normalem]{ulem}
\makeatletter
\def\cyanuwave{\bgroup \markoverwith{\lower3.5\p@\hbox{\sixly \textcolor{cyan}{\char58}}}\ULon}
\def\reduwave{\bgroup \markoverwith{\lower3.5\p@\hbox{\sixly \textcolor{red}{\char58}}}\ULon}
\def\blueuwave{\bgroup \markoverwith{\lower3.5\p@\hbox{\sixly \textcolor{blue}{\char58}}}\ULon}
\font\sixly=lasy6 
\makeatother

\def\BibTeX{{\rm B\kern-.05em{\sc i\kern-.025em b}\kern-.08em
    T\kern-.1667em\lower.7ex\hbox{E}\kern-.125emX}}
\newif\ifdraft{}

\ifdraft{}
  \newcommand{\amnote}[1]{ \textcolor{blue} { ***andrem: #1 }}
  \newcommand{\jhanote}[1]{ {\textcolor{red} { ***shantenu: #1 }}}
  \newcommand{\mtnote}[1]{ {\textcolor{orange} { ***matteo: #1 }}}
\else
  \newcommand{\amnote}[1]{}
  \newcommand{\jhanote}[1]{}
  \newcommand{\mtnote}[1]{}
\fi

\lstdefinestyle{myListing}{
  frame=single,   
  backgroundcolor=\color{listinggray},  
  language=C,       
  basicstyle=\ttfamily \footnotesize,
  breakautoindent=true,
  breaklines=true
  tabsize=2,
  captionpos=b,  
  aboveskip=0em,
  belowskip=-2em,
}      

\lstdefinestyle{myPythonListing}{
  frame=single,   
  backgroundcolor=\color{listinggray},  
  language=Python,       
  basicstyle=\ttfamily \footnotesize,
  breakautoindent=true,
  breaklines=true
  tabsize=2,
  captionpos=b,  
}

\ifpdf{}
  \DeclareGraphicsExtensions{.pdf, .jpg, .tif}
\else
  \DeclareGraphicsExtensions{.eps, .jpg, .ps}
\fi

\journal{SoftwareX}

\begin{document}
\begin{frontmatter}

\title{RADICAL-Cybertools: Middleware Building Blocks for Scalable Science}

\author{Vivek Balasubramanian, Shantenu Jha, Andre Merzky and Matteo Turilli}
\address{Electrical \& Computer Engineering, Rutgers University, Piscataway,NJ 08854, USA}

\begin{abstract}

RADICAL-Cybertools (RCT) are a set of software systems that serve as
middleware to develop efficient and effective tools for scientific computing.
Specifically, RCT enable executing many-task applications at extreme scale
and on a variety of computing infrastructures. RCT are building blocks,
designed to work as stand-alone systems, integrated among themselves or
integrated with third-party systems. RCT enables innovative science in
multiple domains, including but not limited to biophysics, climate science
and particle physics, consuming hundreds of millions of core hours. This
paper provides an overview of RCT components, their impact, and the
architectural principle and software engineering underlying RCT\@.
\end{abstract}


\end{frontmatter}



\section{Motivation and significance}\label{sec:motivation}


The design of distributed systems to support scientific computing has never
been more challenging. Unprecedented diversity in application requirements,
and disruptive changes in the resources and technology landscapes, intermix
with new discovery modalities and need for scalable computing.

Set against this dynamic landscape, two critical question must be addressed:
How can middleware be designed and implemented to meet the collective
challenges of scale, new and diverse functionality, and usability? How can
critical middleware components be designed to be sustainable software
implementations while being forward looking and enable innovative
capabilities?

RADICAL-Cybertools (RCT) are a set of systems developed to address these
challenges.  RCT are building blocks, which can be used as a stand-alone
system, or integrated with other RCT, or third-party tools to enable diverse
functionalities. RCT offer several innovative features to support the design
and implementation of middleware.

This paper takes a software perspective to present the overarching
architectural paradigm of RCT, discussing the design and implementation of
two cybertools: RADICAL-Pilot and Ensemble-Toolkit (EnTK). We outline the
direct impact that RCT are having on domain sciences, focusing the discussion
on architectural and design paradigms of middleware for scientific computing.
In this way, we show how RCT further the ``state-of-theory'' and practice of
scientific computing.

\section{Software description}\label{sec:description}


The RADICAL Cyberinfrastructure tools (RCT)~\cite{github-rct} have three main
components: RADICAL-SAGA (RS)~\cite{merzky2015saga}, RADICAL-Pilot
(RP)~\cite{merzky2018using} and RADICAL-Ensemble Toolkit
(EnTK)~\cite{balasubramanian2018harnessing}.

RS is a Python implementation of the Open Grid Forum SAGA standard 
GFD.90~\cite{goodale2006saga}, a high-level interface to distributed
infrastructure components like job schedulers, file transfer and resource
provisioning services. RS enables interoperability across heterogeneous
distributed infrastructures, improving on their usability and enhancing the
sustainability of services and tools.

RP is a Python implementation of the pilot paradigm and architectural
pattern~\cite{turilli2018comprehensive}. Pilot systems enable users to submit
pilot jobs to computing infrastructures and then use the resources acquired
by the pilot to execute one or more tasks. These tasks are directly scheduled
via the pilot, without having to queue in the infrastructure's batch system.
RP focuses on High Performance Computing (HPC) resources, enabling the
concurrent and consecutive execution of heterogeneous workloads comprised of
one or more scalar, MPI, OpenMP, multi-process, and multi-threaded tasks.
These tasks can be executed on CPUs, GPUs and other accelerators, on the same
pilot or across multiple pilots.

EnTK supports the concurrent or sequential execution of tasks that can be in
an arbitrary priority relation (i.e., ensemble or pipelines of tasks). EnTK
promotes ensembles of tasks to a high-level abstraction, providing a
programming interface and execution model specific to ensemble-based
applications. EnTK is engineered for scale and a diversity of computing
platforms and runtime systems, agnostic of the size, type and coupling of the
tasks comprising the ensemble.

RCT are designed to work both individually and as an integrated system, with
or without third party systems. This requires a ``Building Block'' approach
to their design and development, based on applying the traditional notions of
modularity at system level. The Building Block approach derives from the work
on Service-oriented Architecture and its Microservice variants, and the
component-based software development approaches where computational and
compositional elements are explicitly
separated~\cite{batory1992design,garlan1995architectural,lenz1988software,clemens1998component,schneider2000components}.
AirFlow, Oozie, Azkaban, Spark Streaming, Storm, or Kafka are examples of
tools that have a design consistent with the building blocks approach and
that have been integrated with RCT~\cite{turilli2019middleware}.


\subsection{Software Architecture}\label{ssec:architecture}


All RCT are stand-alone, distributed systems. Architecturally, each tool
consists of one or more subsystems, each with several components. Components
are isolated into individual processes and some components are used only in
specific deployment scenarios, depending on both application requirements and
resource capabilities. Components are stateless and some of them can be
instantiated concurrently to simultaneously manage multiple entities, like
workflows, workloads, tasks or pilots. This enables scaling of throughput and
tolerance to component failure.

Concurrent components are coordinated via a dedicated communication mesh,
which introduces runtime and infrastructure-specific overheads, but improves
overall scalability of the system and lowers component complexity. Components
can have different implementations; configuration files can tailor each RCT
to specific resources types, workloads, or scaling requirements. Components
exchange data about the entities specific to each RCT and data about the
state of the components and subsystems. Each type of data has dedicated
modules and communication channel, separating communication from coordination
with explicit states and events for each entity.

Ref.~\cite{merzky2015saga} details RADICAL-SAGA architecture and
capabilities. In the rest of the paper, we focus on RP and EnTK, first
introducing each system individually, then showing how RCT as a whole can be
composed to serve diverse use cases.

\subsubsection{RADICAL-Pilot}\label{sssec:arch_rp}

RP implements two main abstractions: Pilot and Compute Unit (CU). Pilots and
CUs abstract away specificities of resources and workloads, making it
possible to schedule workloads either concurrently or sequentially on
resource placeholders. Pilots are such placeholders for computing resources,
where resources are represented independent from architecture and topological
details. CUs are units of work (i.e., tasks), specified as an application
executable alongside its resource and execution environment requirements.
Note that a task is not a function, method, thread or process but a program
that runs as a self-contained executable.

\begin{figure}
    \centering
    \subfloat[ ]{{\includegraphics[width=0.555\textwidth]{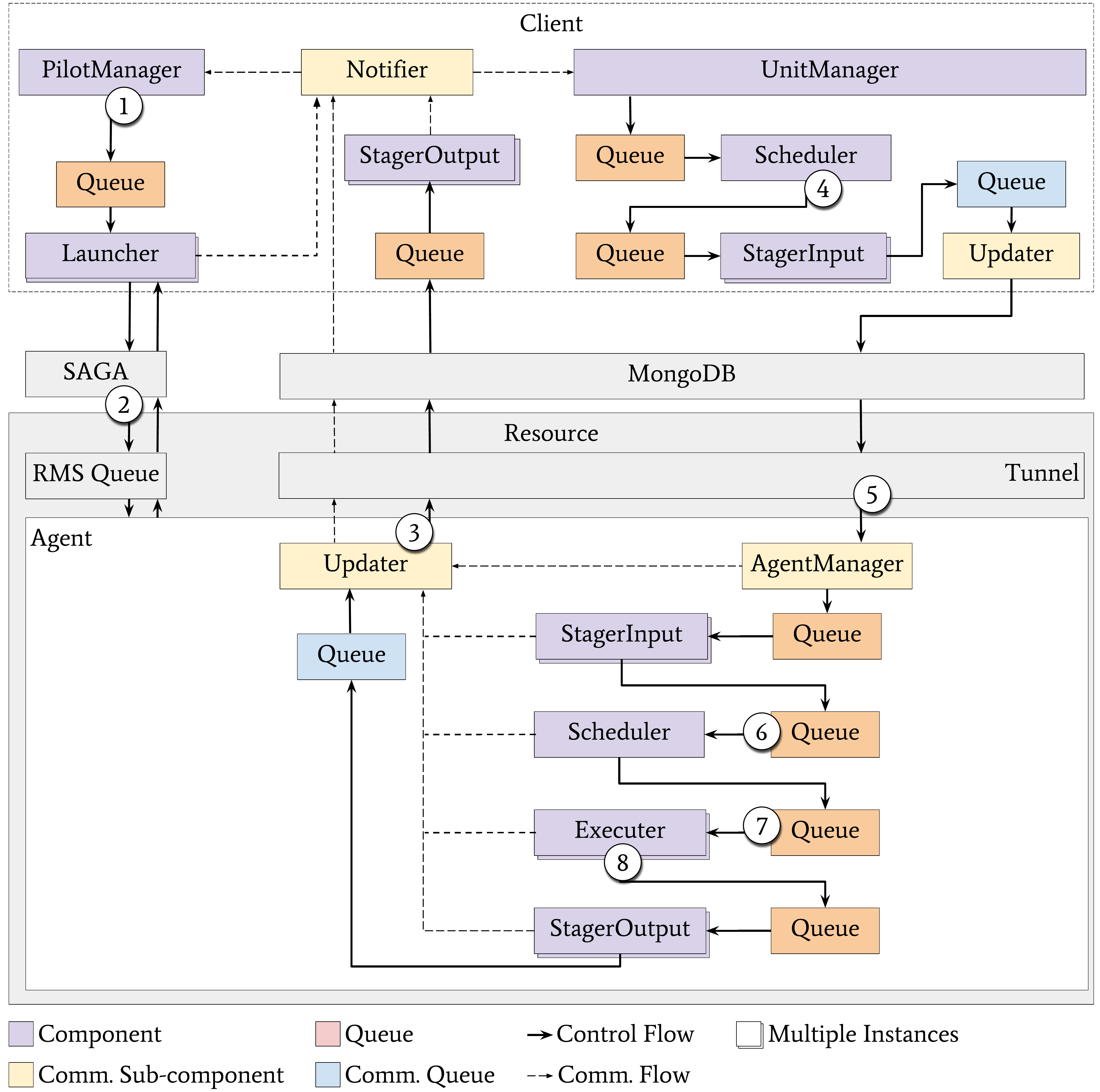} }}
    \qquad
    \subfloat[ ]{{\includegraphics[width=0.34\textwidth]{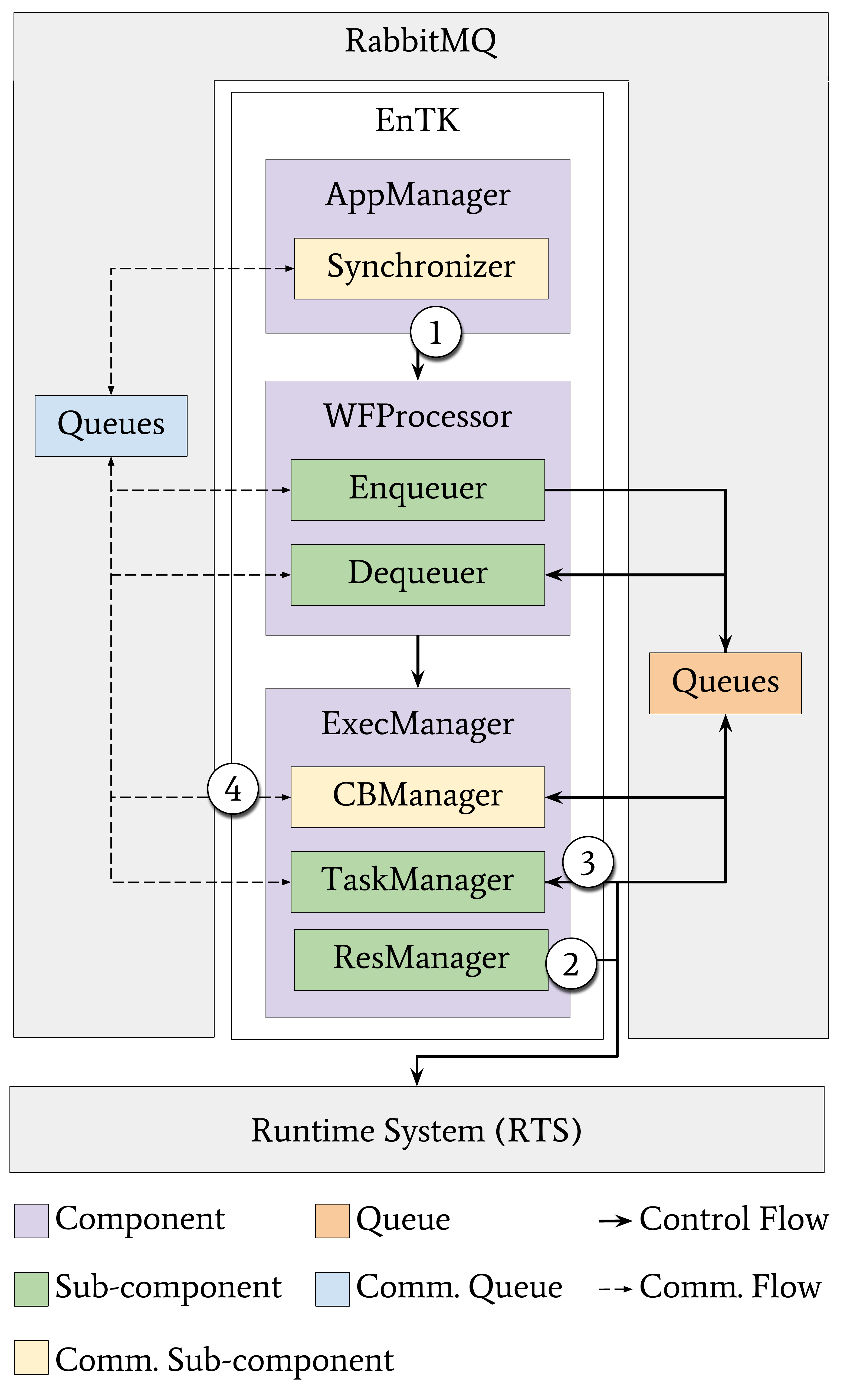} }}
    \caption{Caption. \mtnote{TODO\@: Fix EnTK AppManager}}\label{fig:archs}
\end{figure}

Fig.~\ref{fig:archs}a depicts RP's architecture with two subsystems (white
boxes) and several components (purple and yellow boxes). In each subsystem,
purple components manage pilots and CUs while yellow components manage the
communication among components. Subsystems can execute locally or remotely,
communicating and coordinating over TCP/IP, and enabling multiple deployment
scenarios. For example, users can run Client locally, and distribute MongoDB
and one or more instances of Agent on remote computing infrastructures.
Alternatively, users can run all components on a local or remote resource.

The first subsystem, called Client, has two main components: PilotManager and
UnitManager. PilotManager manages pilots and has a main component called
`Launcher'. Launcher uses resource configuration files to define the number,
placement, and properties of the Agent's components of each pilot. Currently,
configuration files are made available for all the HPC machines of the
Extreme Science and Engineering Discovery Environment (XSEDE), Blue Waters at
the National Center For supercomputing Applications (NCSA), Cheyenne at
NCAR-Wyoming Supercomputing Center (NWSC), and Rhea, Titan and Summit at the
Oak Ridge National Laboratory (ORNL). Users can provide new files or alter
existing configuration parameters at runtime, both for a single pilot or a
whole RP session.

UnitManager manages CUs and has two main components: Scheduler and
StagerInput. Scheduler schedules CUs onto one or more pilots, available on
one or more machines. This enables late binding of CUs to resources,
depending on their availability. CUs are bound to resources that satisfy
their execution requirements only when these resources are actually
available. StagerInput distributes the input files that CU may need for their
execution to the machines on which each CU has been scheduled.

The second subsystem of RP is called `Agent' and has four main components:
StagerInput and StagerOutput for staging CUs' input and output data,
Scheduler and Executer to schedule CUs on a pilot resources and execute those
CUs on them. Multiple instances of the Stager and Executer components can
coexist in a single Agent. Depending on the architecture of the target
machine, Agent's components can individually be placed on cluster head nodes,
MOM/batch nodes, compute nodes, virtual machines, or any combination thereof.
ZeroMQ communication bridges connect the Agent components, creating a network
to support the transitions of CUs through components.

Data management in RP focuses on providing input files to CUs before their
execution, and on moving output files to later CUs, or back into the user
environment. On HPC resources which provide both local and network storage,
RP can select the most appropriate storage, depending on CUs I/O
requirements.  Data can be exclusive to CUs, or can be shared between CUs.

Each component of each subsystem of RP has a dedicated queue to feed entities
into that component. Orange Queues in Fig.~\ref{fig:archs}a are dedicated to
pilots and CUs, blue Queues to the messages exchanged by communication
components. All Queues support bulk communication to obtain performance at
scale in a distributed systems. Further, queues enable load balancing among
concurrent components. Note that concurrent components are used for
performance optimization at scale.

A special queue instance is rendered as collection in a MongoDB database.
That collection is used to communicate between Client and Agent, while
preserving the semantics used for all other queues in Fig.~\ref{fig:archs}a.
Since the MongoDB entries are persistent, that database is also used to store
data for \textit{post mortem} profiling and analysis.

\subsubsection{RADICAL Ensemble Toolkit}\label{sssec:arch_entk}

EnTK implements three main abstractions: task, stage and pipeline. Tasks
contain information regarding an executable, its software environment and its
data dependences. Stages are a set of tasks where the tasks have no mutual
dependences and can therefore execute concurrently, depending on resource
availability. Pipelines are lists of stages where a stage \(i\) can be
executed only after stage \(i-1\) has completed.


As for RP, a task is an executable, i.e., a program. This is important: EnTK
enables concurrent and sequential execution at program-level, not at function
or method level. Parallelism is still possible within each program run by an
EnTK task, enabling concurrent execution of multi-threaded, multi-process and
MPI programs. Note that, at the moment, EnTK requires a runtime system (RTS)
that support the same task abstraction for task execution.

Fig.~\ref{fig:archs}b shows the architecture of EnTK (white) with three
components (purple), each with subcomponents dedicated to the management of
EnTK's entities (green) or coordination of the entities' execution (yellow).
The three components are AppManager, WFProcessor and TaskManager that enable
workflow specification, workflow execution management, and workload
management.

AppManager exposes an API for the development of ensemble-based applications
in terms of tasks, stages and pipelines, and for specifying resource
requirements for the application execution. AppManager initializes EnTK and
holds the global state of the application at runtime. AppManager is the sole
stateful component of EnTK, allowing to restart other components upon
failure, without interrupting the execution of the ensemble-based
application.

WFProcessor uses the Enqueue and Dequeue subcomponents to queue and dequeue
tasks pulled from AppManager. TaskManager uses ExecManager to schedule tasks
on the RTS and keep track of the state of each task during execution.
TaskManager uses ResManageer as an interface to the chosen RTS. RTS have to
provide capabilities to acquire resources and schedule tasks on those
resources for execution. ResManager isolates RTS from EnTK, enabling
restarting of the RTS without loosing information about tasks that have been
already executed. Currently, EnTK support only RP as RTS but it is designed
to use other task-based RTS as, for example, Coasters or HTCondor.

\subsection{Software Functionalities}\label{ssec:functionalities}



As a pilot system, the defining capability of a RP is to decouple resource
acquisition from task execution, enabling execution of tasks on a pilot
without using the resource's queuing system. RP offers an API to describe
both pilots and CUs, alongside classes and methods to manage acquisition of
resources, scheduling and execution of CUs on those resources, and the
staging of input and output files. Reporting capabilities and notifications
update the user about ongoing executions, and profiling capabilities enable
detailed postmortem analysis of workload executions and runtime behavior.



RP offers four unique features when compared to other pilot systems or tools
that enable the execution of many-task applications on HPC systems: (1)
concurrent execution of heterogeneous tasks on the same pilot; (2) support of
all the major HPC batch systems; (3) support of more than twelve methods to
launch tasks; and (4) a general purpose architecture. RP can execute single
or multi core tasks within a single compute node, or across multiple nodes,
isolating the execution of each tasks into a dedicated process and enabling
concurrent execution of heterogeneous tasks by design.


As a workflow engine, EnTK is designed to execute ensemble applications,
respecting the relations of priority among tasks. Compared to similar
systems, EnTK allows to codify relations in terms of pipelines, stages and
tasks, where relations may be determined by input/output data or control flow
requirements. For example, two tasks may have to be executed sequentially
when the output of the first task is the input of the second task; and two
tasks (or ensembles) may have to be executed sequentially when the output of
the first tasks determines whether executing the second task.

Consistently, EnTK provides adaptive capabilities and dedicated constructs to
pause, resume and stop pipelines at runtime. Adaptive applications change the
ensemble specifications, creating new pipelines, stages and tasks, or
changing the properties of those already defined. Further, pipelines and
stages can be paused while waiting to perform \textit{ad hoc} computations.
This enables the implementation of high-level application patterns as, for
example, simulation-analysis or replica exchange.




\section{Illustrative Examples}\label{sec:examples}




Multiple scientific domains can benefit from executing many-task applications
at scale, especially at the scale enabled by leadership-class HPC
machines~\cite{raicu2008many,iosup2011performance}. Independent of the domain
for which these applications are developed, their execution requires to run a
single task, a bag of tasks, or a workflow. In this context, tasks refers to
programs like, for example, GROMACS, NAMD, AMBER, AthenaMP, SPECFEM and many
others. Many-task applications requires to concurrently run multiple
instances of programs, using scale to reduce the total time to completion of
the whole execution.

As seen in Sec.~\ref{sec:description}, RCT support the execution of a single
task, a bag of tasks, and workflows expressed as a set or a sequence of
pipelines with stages and tasks. Because of the separation between manging
the concurrent and consecutive execution of tasks, and the computation
performance by each task, RCT support many-task application independent from
the scientific domain in which they are used. From RCT point of view, every
execution reduces exclusively to manging the execution of single or multiple
sets of programs in the form of black boxes.

RP executes set of tasks. The degree of concurrency of the execution depends
on the amount of available resources. Consider for example a many-task
application for the simulation of molecular dynamics with an ensemble of 128
GROMACS simulations, each requiring 24 CPU cores as those used in
Ref.~\cite{balasubramanian2016extasy}. The user can use RP API to describe a
pilot job with 3072 cores (Lis.~\ref{code:pilot}), 128 CUs
(Lis.~\ref{code:units}) and two managers to coordinate the acquisition of the
pilot resources via RADICAL-SAGA on an HPC machine and the execution of the
128 tasks on those resources (Lis.~\ref{code:mgrs}).

\lstset{language=Python, 
        frame=lines,
        caption={RADICAL-Pilot API: define a 3072-core pilot that runs for 
                 120 minutes on resource `target'},
        label={code:pilot},
        basicstyle=\footnotesize,
        breaklines=true,
        captionpos=b,
        showspaces=false,
        showstringspaces=false,
        showtabs=false,
        tabsize=2}
\begin{lstlisting}
pdesc = rp.ComputePilotDescription()        
pdesc.resource = target
pdesc.cores = 3072
pdesc.runtime = 120
pdesc.project = config[resource]['project']
pdesc.queue = config[resource]['queue']
pdesc.access_schema = config[resource]['schema']
\end{lstlisting}

\lstset{language=Python, 
        frame=lines,
        caption={RADICAL-Pilot API: define 128 24-cores MPI CU 
                 descriptions.},
        label={code:units},
        basicstyle=\footnotesize,
        breakatwhitespace=true,
        breaklines=true,
        captionpos=b,
        showspaces=false,
        showstringspaces=false,
        showtabs=false,
        tabsize=2}
\begin{lstlisting}
n = 128   # number of units to run
cuds = list()
for i in range(0, n):
    cud = rp.ComputeUnitDescription()
    cud.executable = "/bin/bash"
    cud.pre_exec = ['module load gromacs/5.1.2']
    cud.arguments = ['-l', '-c', "/opt/gromacs/bin/gmx_mpi mdrun -s min.tpr -v -deffnm npt"]
    cud.input_staging = ['FRF.itp', 'dynamic.mdp', 'FF.itp', 'martini_v2.2.itp', '85-20.top', 'init85-20.gro'] 
    cud.cores = 24
    cud.mpi = True    
    cuds.append(cud)
\end{lstlisting}

\lstset{language=Python, 
        frame=lines,
        caption={RADICAL-Pilot API: create pilot and unit managers, submit
                 units, and wait for their completion.},
        label={code:mgrs},
        basicstyle=\footnotesize,
        breaklines=true,
        captionpos=b,
        showspaces=false,
        showstringspaces=false,
        showtabs=false,
        tabsize=2}
\begin{lstlisting}
pmgr = rp.PilotManager(session=session)
pilot = pmgr.submit_pilots(pdesc)
umgr = rp.UnitManager(session=session)
umgr.add_pilots(pilot)
umgr.submit_units(cuds)
umgr.wait_units()
\end{lstlisting}

Fig.~\ref{fig:archs}a's numbers illustrate the resource acquisition and task
execution process. PilotManager queues the pilot description on one of the
available Launcher in RP client (Fig.~\ref{fig:archs}a.1). That Launcher uses
RADICAL-SAGA to schedule the pilot as a job on the target resource via the
resource's batch System (Fig.~\ref{fig:archs}a.2). The pilot job waits in the
resource management system queue and, once scheduled, bootstraps the pilot's
AgentMnager and Updater. AgentManager forks the StagerInput, Scheduler,
Executor and StagerOutput components and the Updater notifies RP Client's
Notifier that RP Agent is ready to execute tasks (Fig.~\ref{fig:archs}a.3).

Upon notification, UnitManager queues all the available tasks onto Client's
Scheduler that, in turns, queues those tasks into the StagerInput, depending
on the chosen scheduling algorithm (Fig.~\ref{fig:archs}a.4). If required,
StagerInput stages the tasks' input files to the target resource and then
tasks are queued to the Updater and passed to the chosen RP Agent's
AgentManager (Fig.~\ref{fig:archs}a.5). At that point, tasks are passed to a
StagerInput where input files are linked and made available to each task, and
then queued to the RP Agent's Scheduler (Fig.~\ref{fig:archs}a.6). Scheduler
places tasks on suitable partitions of the pilot's resources and then queues
tasks to the Executor so that, when those partitions of resource becomes
available, tasks are executed (Fig.~\ref{fig:archs}a.7). Executor sets up the
environment required by each task and then forks each task for execution
(Fig.~\ref{fig:archs}a.8). This is why tasks are black boxes to RP\@; also
note that Scheduler and Executor can place and fork heterogeneous tasks,
i.e., task requiring different type and amount core/GPUs and different
execution time.

RP API cannot express dependences among tasks. For RP, every task that is
passed to UnitManager is assumed to be ready for execution. For example,
assume a typical simulation-analysis workflow for molecular dynamics with a
simulation stage and an analysis stage that depends upon the completion of
the simulation stage. Users can explicitly code priorities among stages in
the applications they write with the RP API but they have no dedicated
abstractions in that API for expressing those priorities. EnTK offers these
abstractions at API level: each stage of each pipeline is submitted to RP for
execution, respecting their priority relation.

Fig.~\ref{fig:archs}b's numbers illustrate the execution of workflows in
EnTK\@. Users instantiate an AppManager (Lis.~\ref{code:amgr}), define a set
of resources on which to run their workflow (Lis.~\ref{code:res}), describe
that workflow in terms of pipelines, stages and tasks (Code~\ref{code:pst})
and execute it (Code~\ref{code:exec}). AppManager passes a copy of the
workflow description to WFProcessor that, based on the priorities between
stages and tasks, uses Enquerer to queue tasks that are ready for execution
to the task manager (Fig.~\ref{fig:archs}b.1). Meanwhile, ResManager users
the chosen runtime system to acquire the requested resources
(Fig.~\ref{fig:archs}b.2) and, once available, TaskManagers uses those
resources to execute the queued tasks (Fig.~\ref{fig:archs}b.3) and dequeuing
them once they have been executed. ExecManager uses queues to communicate the
state of each task execution to AppManager (Fig.~\ref{fig:archs}b.4). Note
that AppManager is the only stateful component of EnTK\@: both WFProcessor
and ExecManager can fail without loss of information about the execution.

\lstset{language=Python, 
        frame=lines,
        caption={RADICAL-EnsembleToolkit (EnTK) API: Create AppManager.},
        label={code:amgr},
        basicstyle=\footnotesize,
        breaklines=true,
        captionpos=b,
        showspaces=false,
        showstringspaces=false,
        showtabs=false,
        tabsize=2}
\begin{lstlisting}
appman = AppManager(hostname=hostname, port=port)
\end{lstlisting}

\lstset{language=Python, 
        frame=lines,
        caption={RADICAL-EnsembleToolkit (EnTK) API: Describe a resource request. },
        label={code:res},
        basicstyle=\footnotesize,
        breaklines=true,
        captionpos=b,
        showspaces=false,
        showstringspaces=false,
        showtabs=false,
        tabsize=2}
\begin{lstlisting}
res_dict = {
    'resource': 'target',
    'walltime': 120,
    'cpus': 3072
}
\end{lstlisting}

\lstset{language=Python, 
        frame=lines,
        caption={RADICAL-EnsembleToolkit (EnTK) API: Describe a pipeline with 1 stage with 128 24-cores MPI tasks.},
        label={code:pst},
        basicstyle=\footnotesize,
        breaklines=true,
        captionpos=b,
        showspaces=false,
        showstringspaces=false,
        showtabs=false,
        tabsize=2}
\begin{lstlisting}
p = Pipeline()
s = Stage()
n = 128
for i in range(0, n):
    t = Task()
    t.executable = "/bin/bash"
    t.pre_exec = ['module load gromacs/5.1.2']
    t.arguments = ['-l', '-c', "/opt/gromacs/bin/gmx_mpi mdrun -s min.tpr -v -deffnm npt"]
    t.copy_input_data = ['FRF.itp', 'dynamic.mdp', 'FF.itp', 'martini_v2.2.itp', '85-20.top', 'init85-20.gro'] 
    t.cpu_reqs = {
      'processes': 24,
      'process_type': MPI
      }
    s.add_tasks(t)
p.add_stages(s)
\end{lstlisting}

\lstset{language=Python, 
        frame=lines,
        caption={RADICAL-EnsembleToolkit (EnTK) API: Describe a pipeline with },
        label={code:exec},
        basicstyle=\footnotesize,
        breaklines=true,
        captionpos=b,
        showspaces=false,
        showstringspaces=false,
        showtabs=false,
        tabsize=2}
\begin{lstlisting}
appman.resource_desc = res_dict
appman.workflow = set([p])
appman.run()
\end{lstlisting}

\section{Impact}\label{sec:impact}



The impact of RCT spans domain science, high-performance computing and the
design of software systems. RCT have enabled domain-scientists to achieve
scientific results that would not have been possible otherwise; they have
facilitated research advances in high-performance and distributed computing
systems, while serving as an leading and important prototype implementation
for exploring a paradigmatic shift in the design of middleware for
high-performance scientific workflows.

RCT has enabled the development of scientific applications in multiple and
diverse domains, including software engineering, chemical physics, materials
science, health science, climate science, drug discovery and particle
physics. These users form a worldwide community of domain scientists and
system engineers that actively contribute to the open source development of
RCT. A comprehensive assessment across multiple dimensions is needed to
evaluate the true impact of a software system such as RCT. Whereas the
absolute number of users is a useful metric, an equally important metric is
what those users were able to achieve scientifically and how RCT enabled
them.

Currently, RCT supports a dozen active science projects across the USA and
Europe. The size of projects varies from single PIs with large allocations,
to very large international collaborations. Thus, there is intrinsic
uncertainty in the number of users at any given instant of time but good
faith, best estimates suggest upwards of 30 direct users.

RADICAL-SAGA and RADICAL-Pilot support use cases, spanning functional and
scientific domains. RADICAL-SAGA is mostly integrated into end-to-end
middleware solutions while RADICAL-Pilot is used both as standalone system
and integrated with other systems. As seen in Sec.~\ref{ssec:architecture},
RADICAL-PILOT uses RADICAL-SAGA to submit pilots to a large array of
resources, including HPC and distributed systems.

RADICAL-SAGA enables the Production ANd Distributed Analysis (PanDA) system
to submit batch jobs to Titan and Summit, the two leadership class machines
managed by the Oak Ridge Leadership Computing Facility (OLCF) at the Oak
Ridge National Laboratory (ORNL)~\cite{web-olcf-resources}. PanDA is the
workload management system used by the ATLAS experiment to execute hundred of
millions of jobs a year on both grid and High Performance Computing (HPC)
infrastructures~\cite{maeno2008panda}. The usage of ORNL resources
constitutes 10-12\% of all of ATLAS computing. There are several thousand
researchers that directly or indirectly use PanDA, and thereby RADICAL-SAGA.
In the near future, RADICAL-Pilot will also become a staple of the PanDA
workload management system on HPC platforms.

Reflecting the state of distributed computing systems -- the lack of
simplified and uniform interface to heterogeneous systems, RADICAL-SAGA was
used to develop Science Gateways as part of the Distributed Application
Runtime Environment (DARE) framework. These gateways supported several
projects, including DECIDE and neuGRID, to study the early diagnosis of
Alzheimer and other neurodegenerative diseases. In that capacity RADICAL-SAGA
enabled submission of jobs to distributed computing infrastructures managed
by the European Grid Initiative (EGI), interconnected via GEANT, the
pan-European research and education network that interconnects Europe’s
National Research and Education Networks. Recently, the emergence of toolkits
such as Agave which integrate identity management have provided higher-level
solutions for Gateway developers, but they retain the RADICAL-SAGA based
approach to job submission to distributed computing systems.

Since its first release in 2013, RP has supported a total of two dozen
projects and around 100 active and direct users. Of these, approximately a
dozen projects used RP has a standalone system to support the execution of
many-task applications on single and/or multiple computing infrastructures.
Motivated by the practical lessons from supporting many independent
applications usage of RP as a standalone system, and the realization that an
increasing number of HPC applications were adopting the ensemble
computational model to overcome limitations of single task applications to
achieve significant performance gains on large-scale parallel machines, in
2015 we designed and implemented the Ensemble Toolkit
(EnTK)~\citep{balasubramanian2016extasy} as the latest addition to RCT.

EnTK has enabled the development of domain specific workflow (DSW) frameworks
which provide a specific higher-level functionality. Although, driven by
specific application needs, each DSW is characterized by a unique execution
and coordination pattern and can serve multiple applications. The four
ensemble-based DSW developed using EnTK and other RCT are:
EXTASY~\cite{balasubramanian2016extasy}, RepEx~\cite{treikalis2016repex},
HTBAC~\cite{dakka2018high}, and ICEBERG\@. Details can be found in
Ref.~\citep{turilli2019middleware}

ExTASY and RepEx implement advanced sampling algorithms using biomolecular
simulations. Both use the EnTK API to implement  diverse coordination
patterns amongst ensembles of biomolecular simulations and analysis. HTBAC
supports multiple algorithms that compute free-energy calculations that are
critical to drug design and resistance studies. HTBAC allows the runtime
adaptation at multiple levels: algorithms, pipelines and tasks within a
pipeline. This capability has been demonstrated to reduce the
time-to-solution by a factor 2.5 in controlled experiments on real drug
candidates~\citep{dakka2018concurrent}. ICEBERG supports scalable image
analysis applications using multiple concurrent pipelines.

ExTASY, RepEx, HTBAC and ICEBERG benefit from integrating RCT by not having
to re-implement workflow processing, efficient task management and
interoperable task execution capabilities on distinct and heterogeneous
platforms. This, in turn, enables both a focus on and ease of ``last mile
customization'' for the DSW\@.

RCT are a testbed for engineering research, mostly focused on foundational
abstractions~\cite{turilli2017evaluating}, architectural
paradigms~\cite{turilli2018comprehensive}, application
patterns~\cite{balasubramanian2016extasy,balasubramanian2018harnessing}, and
performance analysis of distributed middleware on diverse computing
infrastructures~\cite{turilli2017evaluating,dakka2018high}. Among the most
representative projects supported by RP as a standalone system, the
Abstractions and Integrated Middleware for Extreme-Scale Science (AIMES)
project enabled extreme-scale distributed computing via dynamic federation of
heterogeneous computing infrastructures. We used RP to execute millions of
tasks on both HPC and HTC resources, studying the federated behavior of
multiple infrastructures, establishing for the first time ever the importance
of integrating task and resource information in scheduling and placement
decision making for federated supercomputers~\cite{turilli2016integrating}.

The Building Blocks approach helps to create systems that can support use
cases both individually and as integrated, end-to-end
solution~\cite{turilli2019middleware}. This is important when supporting
projects with multiple, distinct use cases. For example, ICEBERG has to
support five use cases, each investigating a specific problem in the domain
of polar science. Four of these use cases require the concurrent execution of
pipelines but one requires only the execution of bag of tasks. The first four
cases can use EnTK while the latter only RADICAL-Pilot. Importantly, all use
cases can be served by the ICEBERG framework with a minor change in the
private API to call EnTK or RP, depending on the use cases and therefore
without any engineering overheads.

\section{Conclusions}\label{sec:conclusions}

RCT is a small operation with at most two developers working on the systems
at the same time. In order to implement the aforementioned capabilities while
supporting more than ten concurrent projects at every point in time, we
adopted a specific methodology for the design, development and maintenance of
RCT\@. This methodology is based on the Building Blocks approach, the use of
git for distributed version control of the code base~\cite{github-rct}, and a
tailored project management process.

Overall, these processes and insight have an impact on how to approach the
development of middleware for supporting scientific research. Based on more
than ten years of experience, our approach show a sustainable and effective
way to organize software development, promote community adoption and leverage
the specific characteristics of the academic financial model. This signs the
transition from a development model based on end-to-end, monolithic solutions
with stringent requirements on infrastructures' software stack, to a model
based on small, independent and composable systems, each with a well-defined
capability. Note that these systems must be composable with third-party
systems, i.e., systems developed independently by different development
teams. This approach enables a model of sustainability based on smaller and
shorter funding sources but requires a certain convergence in the vision of
diverse groups competing in the same research field.

\section*{Acknowledgments and Contributions}

We thank all our users, collaborators, and present and past members of
RADICAL\@. This paper was was supported primarily by NSF 1440677 and DOE ASCR
DE-SC0016280. We acknowledge access to computational facilities: XSEDE
resources (TG-MCB090174), Blue Waters (NSF-1713749), and DOE leadership
machines via multiple INCITE and DD awards. In alphabetical order: Vivek
Balasubramanian is the lead developer of EnTK\@; Shantenu Jha is the PI of
RADICAL lab; Andre Merzky is the lead developer and architect of
RADICAL-Pilot and RADICAL-SAGA\@; and Matteo Turilli is the lead architect of
EnTK, co-architect of RADICAL-Pilot and wrote the paper.

\bibliographystyle{elsarticle-num} 
\bibliography{rct-softwarex}

\appendix

\section*{Required Metadata}\label{sec:metadata}

\section*{Current code version}\label{sec:src_version}


\begin{table}[!ht]
\begin{tabular}{|l|p{6.5cm}|p{6.5cm}|}
\hline
\textbf{Nr.}                                                     & 
\textbf{Code metadata description}                               & 
\textbf{Please fill in this column}                              \\
\hline
C1                                                               & 
Current code version                                             & 
0.60                                                             \\
\hline
C2                                                               & 
Permanent link to code/repository used for this code version     & 
\url{https://github.com/radical-cybertools}                      \\
\hline
C3                                                               & 
Legal Code License                                               & 
MIT License (MIT)                                                \\
\hline
C4                                                               & 
Code versioning system used                                      & 
git                                                              \\
\hline
C5                                                               & 
Software code languages, tools, and services used                & 
python, shell, C                                                 \\
\hline
C6                                                               & 
Compilation requirements, operating environments \& dependencies & 
virtualenv, pip or conda                                         \\
\hline
C7                                                               & 
Developer documentation/manual                                   & 
\url{https://radicalpilot.readthedocs.io/}                       \\
\hline
C8                                                               & 
Support email for questions                                      & 
\url{radical-cybertools@googlegroups.com}                        \\
\hline
\end{tabular}
\caption{Code metadata}\label{tab:src_metadata} 
\end{table}

\section*{Current executable software version}\label{sec:bin_version}


\begin{table}[!ht]
\begin{tabular}{|l|p{6.5cm}|p{6.5cm}|}
\hline
\textbf{Nr.}                                                     & 
\textbf{Exec. metadata description}                          & 
\textbf{Please fill in this column}                              \\
\hline
S1                                                               & 
Current software version                                         & 
0.60                                                             \\
\hline
S2                                                               & 
Permanent link to executables of this version                    & 
\url{https://pypi.org/project/radical.pilot/}                    \\
\hline
S3                                                               & 
Legal Software License                                           & 
MIT License (MIT)                                                \\
\hline
S4                                                               & 
Computing platforms/Operating Systems                            & 
GNU/Linux operating systems                                      \\
\hline
S5                                                               & 
Installation requirements \& dependencies                        & 
virtualenv, pip or conda                                         \\
\hline
S6                                                               & 
User manual and publications                                     & 
User manual: \url{https://radicalpilot.readthedocs.io/}; 
Publications: Refs~\cite{merzky2015saga,balasubramanian2018harnessing,merzky2018using}                                       \\
\hline
S7                                                               & 
Support email for questions                                      & 
\url{radical-cybertools@googlegroups.com}                        \\
\hline
\end{tabular}
\caption{Software metadata}
\label{tab:bin_metadata} 
\end{table}

\end{document}